\documentclass[pre,twocolumn,aps,floatfix,10pt,superscriptaddress]{revtex4}
\usepackage{amsmath}
\usepackage{amsfonts}
\usepackage{amssymb}
\usepackage{graphicx}
\usepackage{subfigure}

\providecommand{\norm}[1]{\lVert#1\rVert}

\renewcommand{\Im}{\operatorname{Im}}

\newcommand{\be}{\begin{equation}}
\newcommand{\ee}{\end{equation}}
\newcommand{\bea}{\begin{eqnarray}}
\newcommand{\eea}{\end{eqnarray}}
\newcommand{\vq}{\vec{q}}
\newcommand{\px}{\partial_1}
\newcommand{\py}{\partial_2}
\newcommand{\pz}{\partial_3}

\renewcommand{\Im}{\operatorname{Im}}

\begin{document}

\title{Phase-field-crystal models and mechanical equilibrium}

\author{V. Heinonen}
\affiliation{COMP Centre of Excellence at the
Department of Applied Physics,
Aalto University, School of Science,
P.O.Box 11100,
FI-00076 Aalto
Finland}
\email{vili.heinonen@aalto.fi}
\author{C. V. Achim}
\affiliation{Institut f\" ur Theoretische Physik II:
Weiche Materie, Heinrich-Heine-Universit\" at
D\" usseldorf,  D\" usseldorf, Germany}
\author{K. R. Elder}
\affiliation{Department of Physics, Oakland University,
Rochester, Michigan 48309, USA}
\author{S. Buyukdagli}
\affiliation{COMP Centre of Excellence at the
Department of Applied Physics,
Aalto University, School of Science,
P.O.Box 11100,
FI-00076 Aalto
Finland}
\author{T. Ala-Nissila}
\affiliation{COMP Centre of Excellence at the
Department of Applied Physics,
Aalto University, School of Science,
P.O.Box 11100,
FI-00076 Aalto
Finland}
\affiliation{Department of Physics, Brown University,
Providence RI 02912-1843, U.S.A.}

\begin{abstract}
Phase field crystal (PFC) models constitute a field theoretical approach to
solidification, melting and related phenomena at atomic length and diffusive
time scales. One of the advantages of these
models is that they naturally contain elastic excitations associated with
strain in crystalline bodies. However, instabilities that are diffusively
driven towards equilibrium are often orders of magnitude slower than the
dynamics of the elastic excitations, and are thus not included in the standard
PFC model dynamics. We derive a method to isolate the time evolution of the
elastic excitations from the diffusive dynamics in the PFC approach and set
up a two-stage process, in which elastic excitations are equilibrated
separately. This ensures mechanical equilibrium at all times.
We show concrete examples demonstrating the
necessity of the separation of the elastic and diffusive time scales. In the
small deformation limit this approach is shown to agree with the theory of
linear elasticity.
\end{abstract}


\maketitle

\section{Introduction}

Insight into crystal growth and related phenomena is essential for
understanding fundamental material properties and exploiting them in
engineering applications.  Atomistic methods such
as density functional theory (both quantum-mechanical and classical), or
numerical molecular dynamics simulations have provided
a great deal of information about material properties but are
limited to relatively small length and time scales. Thermally driven dynamics such as
annealing of defects or relaxing stress in heteroepitaxial systems requires
microsecond time scales that are beyond such methods.  On large length
scales there are a vast number of macroscopic field theories
for studying melting and solidification but these theories often fail to
incorporate atomistic details that
are essential for understanding the phenomena associated with crystal growth.
To this end the phase
field crystal (PFC) model was proposed by Elder {\it et al.}
\cite{Elder:2002eq,Elder:2004ct,Elder:2007kl} to add a richer theory
able to describe the underlying
crystalline structure with elastic and plastic properties.
The PFC model describes the dynamics of a dimensionless field $n$ that
is related to the atomic number density and as such is periodic in a crystalline
state and constant in a liquid phase.  PFC models have
been applied to study a wide range of different phenomena such as
grain-boundary melting \cite{Berry:2008ig,Mellenthin:2008hs}
and energy \cite{Jaatinen:2009gb}, fractal growth
\cite{Tegze:2011fo}, surface ordering
\cite{Achim:2006fe,Achim:2009ko,Ramos:2010fa,Ramos:2008jo},
epitaxial growth \cite{Elder:2002eq,Elder:2004ct,Elder:2007kl,Wu:2009},
the yield stress of polycrystals \cite{Stefanovic:2006fg,Hirouchi09,Stefanovic:2009}
and glass transitions \cite{Berry08,Berry11}.

	The dynamics of the PFC model were originally assumed to be conserved
dissipative and driven by the chemical potential to minimize an associated
free energy functional. Such dynamics can in certain limits be justified
by more fundamental arguments \cite{Marconi:1999wb,Archer:2004eq,Sami:2007}.
However, since the energy of the PFC models incorporates elastic energy due to elastic
stress this might turn out to be problematic in some cases.  For example,
elastic excitations such as travelling wave modes and simple stretch or
compression of solid body should relax considerably faster than the diffusive
time scales of solidification. Even in a system that is actively not driven out
of equilibrium elastic excitations can arise. For example, crystallization from
multiple crystallization centres may result into elastic excitations when the
grains that are oriented in different directions meet at the grain boundaries.
It is often justifiably assumed that in many materials
(metals, insulators, semi-conductors) elastic equilibrium is instantaneous
compared to the other slow processes, such as solidification, phase segregation,
etc.  Attempts have been made to address this issue by adding higher
time derivatives to the PFC equation \cite{Stefanovic:2006fg}, but such an approach
is only approximate.  In this work we present a method that can be
shown to give exact elastic or mechanical equilibrium in the small deformation
limit.  This approach relies on the so-called amplitude formulation of the PFC model.

	The amplitude formulation bridges the gap between the conventional PFC model
and more macroscopic phase field models and was introduced by Goldenfeld and
collaborators \cite{Goldenfeld:2005dq,Athreya:2006hs,Goldenfeld:2006by}
for the two dimensional triangular phase of the PFC model and has been
extended to three dimensional bcc and fcc crystals, binary alloys \cite{Elder10,Huang10b}
and to include miscibility gap in the density field \cite{Yeon10}.
This approach
considers variations of the amplitudes of a periodic density field.  The
amplitudes are complex so that they have two degrees of freedom, magnitude
and phase.  Essentially, the magnitude of the amplitude is zero in the
liquid state and finite in a crystalline phase, while the phase can account
for elastic deformations and rotations.  The combination of the two can
describe dislocations and grain boundaries, since the phase can be discontinuous
when the magnitude goes to zero. This approach allows for larger length and time scales,
and as shown by Athera {\it et al.} \cite{Athreya:2007fc} can be numerically
implemented using efficient multi-grid methods.  It is also very useful for
studies in which the crystal orientation is almost the same everywhere (except
near dislocations) as in the case of
heteroepitaxial systems \cite{Huang08,Huang10,Elder12,Elder13}.

In this article we propose a method to isolate and separately equilibrate the elastic
excitations of the system within the framework of the amplitude expansion of
the PFC model and show that in a certain limit this is consistent with
elastic equilibrium.  Numerical verification of the theory is also provided.
The article is organized as follows: Sec.~\ref{sec:phase-field} introduces the
phase-field crystal model and its corresponding amplitude expansion. The separation of the
fast time scales associated with elastic excitations is described in
Sec.~\ref{sec:separation} and later studied in the linear deformation limit in
Sec.~\ref{sec:linear_deformation}. The theory is numerically tested in
Sec.~\ref{sec:numerics} and finally the results are summarized and concluded in
Sec.~\ref{sec:summary}.

\section{Phase-field crystal model} \label{sec:phase-field}
The phase-field crystal model \cite{Elder:2002eq,Elder:2004ct,Elder:2007kl}
is a coarse-grained model that describes the
dynamics of a dimensionless field $n$ that is related to deviations of
the atomic number density from the average number density.  The associated
free energy can be written in dimensionless form as
%
\begin{equation}
\begin{split}
F_{PFC} [n(\vec{r})] &= \int_{\Omega} d\vec{r}\left\lbrace
\frac{\Delta B}{2} n(\vec{r})^2 + B^x \frac{n}{2}(1+\nabla^2)^2 n
\right. \\& \left.
- \frac{\tau}{3} n^3
+ \frac{v}{4}n^4 \vphantom{\frac{\Delta B}{2}}
\right\rbrace,
\end{split}
\label{eq:PFC_energy}
\end{equation}
where $\Delta B \equiv B^\ell-B^x$. The parameter $B^\ell$ is related to the
compressibility of the liquid state, and the elastic moduli of the
crystalline state are proportional to $B^x$. The parameters $\tau$ and
$v$ control the amplitude of the fluctuations in the solid state and
the liquid-solid miscibility gap. Descriptions of these parameters
can be found in Refs. \cite{Jaatinen:2009gb,Elder:2007kl}.
The field $n$ is a conserved quantity that is driven to minimize the
free energy, i.e.,
\begin{equation}
\begin{split}
\frac{d n}{d t} &= \nabla^2 \frac{\delta F_{PFC}}{\delta n} \\ &
= \nabla^2 \left[ \Delta B n + B^x (1+\nabla^2)^2 n - \tau n^2 + v n^3
\right],
\end{split}
\label{eq:pfc_evolution}
\end{equation}
where the mobility has been set to unity.
As discussed in many previous works, the free energy functional has
an elastic contribution and Eq. (\ref{eq:pfc_evolution}) does relax
to minimize any elastic deformations.  However, the relaxation
is on diffusive times scales, not on time scales associated
with phonon modes (or speed of sound time scales) that can be significantly
faster in metals and semiconductors.  It is often assumed that the relaxation
is instantaneous compared to processes such as vacancy diffusion
or solidification in traditional phase field models of such
systems \cite{Muller99,Haataja02,orl99,wang01}.

	To see how instantaneous elastic equilibrium can be
achieved it is useful to consider an amplitude representation
of $n$ and the corresponding equations of motion. In the solid
phase the ground state can be represented in an amplitude expansion, i.e.,
\begin{equation}
n \approx \bar{n} + \sum_{j} \left[
\eta_{j}(\vec{r},t) e^{i \vec{q}_{j} \cdot \vec{r} } + \textrm{c.c.}
\right],
\label{eq:amp}
\end{equation}
where $\eta_{j}$ are complex amplitudes, $\vec{q}_{j}\equiv
{\rm k}\,\vec{k}_1+{\rm l}\,\vec{k}_2+{\rm m}\,\vec{k}_3$, (klm) are the Miller indices
and $(\vec{k}_1,\vec{k}_2,\vec{k}_3$) are the principle reciprocal lattice
vectors.  Within a certain range of parameters,
Eq.~\eqref{eq:PFC_energy} produces a triangular crystal lattice in two
dimensions as a ground state that to a good approximation can be
approximated by only three amplitudes, the ones corresponding to
the three smallest $|\vec{q}_j|$'s.  More generally, equations of motion
for the amplitudes have been derived using various methods by assuming
that they are slowly varying functions of space and times.
This procedure is described in detail in Refs.~\cite{Athreya:2006hs,Goldenfeld:2006by,
Goldenfeld:2005dq,Yeon10,Elder10,Huang10b}.

	The reason it is interesting to consider the complex amplitudes
as opposed to $n$ itself is that the magnitude and phase of $\eta_j$
describe different physical features and more
importantly naturally separate out the elastic part, as will be
explained in the next section.  This separation
makes it possible to relax the elastic energy at a different rate
than other processes, such as vacancy diffusion and climb.  The separation
of time scales is discussed in the next section.

\section{Amplitude expansion}\label{sec:separation}
A simple derivation of the equations of motion of the amplitudes
can be obtained by assuming that the amplitudes are approximately
constant at atomic length scales.  This is done by
substituting  Eq. (\ref{eq:amp})
into Eq. (\ref{eq:pfc_evolution}),
multiplying by $e^{-i\vec{q}_j\cdot\vec{r}}$ and averaging over one
unit cell, assuming that the $\eta_j$'s are constant.  This heuristic
method gives essentially the same result as more
rigorous multiple scales or renormalization group calculations.
The results of such calculations give the
dynamic amplitude equations for a single component
2D system \cite{Elder10},
\begin{equation}
\begin{split}
\frac{d \eta_j}{d t} &=
- \left\lbrace \vphantom{\prod_{i \neq j}}
\left[\Delta B + B^x \mathcal{G}_j^2
+ 3v(A^2-|\eta_j|^2)\right] \eta_j
\right. \\ &\left.
 - 2\tau \prod_{i \neq j} \eta_i^*
\right\rbrace,
\end{split}
\label{aevolution}
\end{equation}
where
\begin{eqnarray}
\mathcal{G}_j &\equiv& \nabla^2 + 2i \vec{q}_j \cdot \vec{\nabla}, \\
A^2 &\equiv& 2 \sum_j |\eta_j|^2, \\
\vec{q}_1 &=&-\sqrt{3} \hat{x}/2-\hat{y}/2,\\
\vec{q}_2 &=&\hat{y},\\
\vec{q}_3 &=&\sqrt{3} \hat{x}/2-\hat{y}/2.
\end{eqnarray}
The time evolution can also be written by using the (dimensionless) free energy as
\begin{equation}
\frac{d \eta_j}{d t} = -\frac{\delta F}{\delta \eta_j^*},
\end{equation}
with the free energy
\begin{equation}
\label{eq:2Denergy}
\begin{split}
F[\{\eta_j \}] &=\int_{\Omega} d\vec{r} \left\lbrace
 \frac{\Delta B}{2} A^2
+\frac{3v}{4} A^4 \right. \\ &\left.
+ \sum_{j=1}^3 \left[
B^x |\mathcal{G}_j \eta_j|^2-\frac{3v}{2}|\eta_j|^4 \right] \right. \\& \left.
-2\tau ( \eta_1 \eta_2 \eta_3 + \eta_1^* \eta_2^* \eta_3^*) \vphantom{\frac{\Delta B}{2}}
\right\rbrace .
\end{split}
\end{equation}
It should be noted that $|\vec{q}_j|=1$, which is a direct consequence of
the term $(1+\nabla^2)^2$ in the PFC energy defined by
Eq.~\eqref{eq:PFC_energy}. This sets the length scale throughout the article.

The complex amplitudes represent renormalized amplitudes of a one-mode
approximation of the number density field associated with the PFC models. The
approximate PFC density field can be reconstructed from Eq. (\ref{eq:amp}).
A perfect crystal can be realized by setting $\eta_j=\phi=\textit{constant}$.
This helps in the interpretation of the complex amplitudes. Consider a
deformation of the coordinates $\vec{r} \to \vec{r}+\vec{u}$ with some
deformation field $\vec{u}(\vec{r})$.  Making this substitution in $n$
is equivalent to transforming the amplitudes of the perfect crystal to
$\eta_j = \phi \exp{(i \vec{q}_j \cdot \vec{u} )}$.  Thus we see that
the phase of the amplitude carries information about the deformation field
$\vec{u}$. For this reason it is useful to write $\eta_j = \phi_j \exp(i\theta_j)$
and consider the equations of motion for $\phi_j$ and $\theta_j$
separately.

Writing  $\eta_j = \phi_j \exp{(i \theta_j)}$ the complex
amplitudes can be separated into fields $\phi_j(\vec{r},t)$ that differentiate
between the liquid and solid phase, and fields $\theta_j(\vec{r},t)$ that represent
deformations. The idea behind the separation of the time scales is that the fields
$\phi_j(\vec{r},t)$ represent slow melting, solidification and diffusive phenomena while the
fields $\theta_j(\vec{r},t)$ stand for deformations that are in general fast.
It is then straightforward to show that Eq. (\ref{aevolution}) becomes
\begin{equation}
\begin{split}
\frac{d \phi_j}{d t} + i \phi_j \frac{d \theta_j}{d t} &=
- B^x (\mathcal{L}_j^2-4\mathcal{Q}_j^2+4i \mathcal{Q}_j \mathcal{L}_j)\phi_j  \\&
-\Delta B \phi_j
-3v \left( 2 \sum_i (\phi_i^2) - \phi_j^2 \right)\phi_j \\&
+ 2\tau \left( \prod_{i\neq j} \phi_i
\right) \exp{\left( -i \sum_i \theta_i \right)},
\end{split}
\label{eq:motions}
\end{equation}
where $\mathcal{Q}_j$ and $\mathcal{L}_j$ are operators given by
\be
\mathcal{Q}_j \equiv \vec{q}_j \cdot \left( \vec{\nabla} + i \vec{\nabla} \theta_j\right),
\ee
and
\be
\mathcal{L}_j \equiv \nabla^2 -|\vec{\nabla}\theta_j|^2+2i\vec{\nabla}\theta_j \cdot
\vec{\nabla} + i \nabla^2\theta_j.
\ee
Collecting the real and the imaginary parts of the right-hand side of
Eq.~\eqref{eq:motions} gives the equations of motion for $\phi_j$ and
$\theta_j$.  To understand the behavior of the fields and their relationship
to elastic equilibrium it is useful to consider next the limit of a small
deformation.

\section{Small deformation limit} \label{sec:linear_deformation}
In the small deformation limit the complex amplitudes
can be represented as $\eta_j = \phi e^{i\vec{q}_j \cdot \vec{u}}$, where
$\vec{u}$ is the standard displacement vector used in continuum elasticity
theory \cite{landau1986theory}.  If we consider the case in which the derivatives of
$\vec{u}$ are of linear order and $\phi$ is constant in time and space, it is then
straightforward to determine the conditions in which
elastic equilibrium is reached.   In the next section these
conditions are derived by
varying the free energy with respect to
the strain tensor
to determine the stress tensor and then imposing
the standard definition of elastic equilibrium, i.e., that
the divergence of the stress tensor is zero.
In subsection \ref{subsec:eqmeq} it is also shown that this is
equivalent to a condition on the dynamics of the phases $\theta_j$.
In this way new equations for the dynamics of the phases can be
introduced to ensure exact elastic equilibrium in the small
deformation limit.

\subsection{Elastic equilibrium from energy}
In the linear elastic limit the elastic equilibrium is written in terms of
the linear strain tensor
\begin{equation}
\varepsilon_{ij} = \frac{1}{2} \left( \frac{\partial u_i}{\partial x_j} +
\frac{\partial u_j}{\partial x_i} \right), \end{equation}
where $u_i$ are the components of the deformation field $\vec{u}$ and the
stress tensor that can be obtained by taking a tensor derivative of
the free energy with respect to the strains, i.e.,
\begin{equation}
\sigma_{ij} =
\begin{cases}
\frac{\delta F}{\delta \varepsilon_{ij} } & \text{if $i=j$}; \\
\frac{1}{2} \frac{\delta F}{\delta \varepsilon_{ij} } & \text{if $i\neq j$}.
\end{cases}
\label{eq:stress}
\end{equation}
Elastic equilibrium is then established when,
\begin{equation}
\vec{\nabla} \cdot \vec{\vec{\sigma}} = 0.
\label{eq:newton2}
\end{equation}
The above equation is Newton's second law of motion for continuum
media in the static case.

	As shown in a prior publication \cite{Elder10}, for a two
dimensional triangular lattice the elastic contribution to the free energy
given in Eq. (\ref{eq:2Denergy}) can be written as

\begin{equation}
\begin{split}
F_{el} &= \int d\vec{r}  \left[
3 B^x \phi^2 \left(
 \frac{3}{2} \varepsilon_{11}^2
 +\frac{3}{2} \varepsilon_{22}^2
+\varepsilon_{11} \varepsilon_{22}
+2 \varepsilon_{12}^2
\right) \right].
\end{split}
\end{equation}
Using this we can write down the components of the stress tensor
from Eq. (\ref{eq:stress})
\begin{equation}
\begin{split}
\sigma_{11}
=3B^x \phi^2 ( 3\varepsilon_{11} + \varepsilon_{22})
=3B^x \phi^2 (3 \partial_1 u_1 + \partial_2 u_2),
\end{split}
\end{equation}
\begin{equation}
\begin{split}
\sigma_{22}
=3B^x \phi^2 ( 3\varepsilon_{22} + \varepsilon_{11})
=3B^x \phi^2 (\partial_1 u_1 + 3 \partial_2 u_2),
\end{split}
\end{equation}
and
\begin{equation}
\begin{split}
\sigma_{12}
=6 B^x \phi^2 \varepsilon_{12}
= 3 B^x \phi^2 (\partial_1 u_2 + \partial_2 u_1).
\end{split}
\end{equation}
Elastic equilibrium follows from using Eq.~\eqref{eq:newton2} as
\begin{equation}
(\vec{\nabla}\cdot\vec{\vec{\sigma}})_1 =
3B^x \phi^2 ( 3\partial_1^2 u_1 + \partial_2^2 u_1+ 2 \partial_1 \partial_2 u_2 )
=0,
\label{eq:m1}
\end{equation}
\begin{equation}
(\vec{\nabla}\cdot\vec{\vec{\sigma}})_2 =
3B^x \phi^2 (3 \partial_2^2 u_2 + \partial_1^2 u_2 + 2 \partial_1 \partial_2 u_1 )
=0,
\label{eq:m2}
\end{equation}
simplifying to
\begin{align}
\label{eq:enemecheq1}
3\partial_1^2 u_1 + \partial_2^2 u_1+ 2 \partial_1 \partial_2 u_2&= 0, \\
\label{eq:enemecheq2}
3 \partial_2^2 u_2 + \partial_1^2 u_2 + 2 \partial_1 \partial_2 u_1 &=0.
\end{align}
These equations describe the elastic equilibrium conditions for triangular
crystal systems, where the linear strain tensor is connected to the stress via
elastic constants as
\begin{align}
&\sigma_{ii} = C_{11} \varepsilon_{ii} + C_{12} \varepsilon_{jj},\quad \textrm{for $i\neq j $}; \\
&\sigma_{12} = \sigma_{21} =2 C_{44} \varepsilon_{12}.
\end{align}
The above calculations give the elastic constants the
values of $C_{11}=9 B^x \phi^2$,
$C_{12}=3B^x \phi^2$, and $C_{44}=3 B^x \phi^2$.

\subsection{Elastic equilibrium from dynamical equations} \label{subsec:eqmeq}
Eq.~\eqref{eq:motions} gives the time evolution for the fields $\phi_j$ and
$\theta_j$.  In the limit that $\phi$ is constant in space and time the
real and imaginary parts of Eq. ~\eqref{eq:motions} become,
\be
4\vec{q}_j \cdot \vec{\nabla} \nabla^2 \theta_j = C,
\label{eq:real}
\ee
and
\be
\frac{d\theta_j}{dt} = - B^x\left(\nabla^4
-4(\vec{q}_j\cdot\vec{\nabla})^2\right)\theta_j,
\label{eq:imag}
\ee
respectively, where $C\equiv \Delta B \phi -2 \tau \phi^2+15v\phi^3$.
In the small deformation limit, $\theta_j \equiv \vec{q}_j \cdot \vec{u}$,
Eq. (\ref{eq:real}) for $j=2$ becomes
\be
4B^x \partial_2 \nabla^2 u_2 = C,
\ee
which implies
\be
\partial_2^2 \nabla^2 u_2 = \partial_1\partial_2 \nabla^2 u_2 = 0.
\label{eq:vy}
\ee
Similarly, by adding Eq. (\ref{eq:real}) for $j=1$ and $j=3$ it is
easy to show that
\be
\partial_1^2 \nabla^2 u_1
= \partial_2\partial_1 \nabla^2 u_1 = 0.
\label{eq:vx}
\ee
In addition, subtracting Eq. (\ref{eq:real}) for $j=1$ and $j=3$ gives
\be
\partial_1 \nabla^2 u_2 = - \partial_2 \nabla^2 u_1.
\ee
By taking derivatives of this equation and using  Eqns. (\ref{eq:vy}) and
(\ref{eq:vx}) it is straightforward to show that
\be
\partial^2_1 \nabla^2 u_2 = \partial^2_2 \nabla^2 u_1 =0
\ee
or $\nabla^4\vec{u}=0$ which implies $\nabla^4\theta_j=0$.  Thus
Eq. (\ref{eq:imag}) to becomes simply
\be
\frac{d\theta_j}{dt} =  4 B^x (\vec{q}_j\cdot\vec{\nabla})^2 \theta_j.
\label{eq:imags}
\ee

In the small deformation limit the deformation field $\vec{u}$ can be written as
\be
\vec{u} = \frac{2}{3}\sum_{j=1}^{3}
\vec{q}_j \theta_j 
\ee
using deformations along reciprocal lattice vectors
$\vec{q}_j$ given by $\theta_j = \vec{q}_j \cdot \vec{u}$.
The condition for elastic equilibrium becomes
\be
\frac{d \vec{u}}{dt} = \frac{2}{3}\sum_{j=1}^{3} \vec{q}_j \frac{d\theta_j}{dt} = 0.
\ee
For a two-dimensional triangular system we have
\begin{equation}
\begin{split}
\sum_{j=1}^{3} q_{j1} \frac{d\theta_j}{dt} &=
\frac{\sqrt{3}}{2}\frac{d\theta_3}{dt}-\frac{\sqrt{3}}{2}\frac{d\theta_1}{dt}  \\
&= \frac{3 B^x}{2} \left(3\partial_1^2 u_1
+ \partial_2^2 u_1+ 2 \partial_1 \partial_2 u_2\right)  \\
&= \frac{1}{2 \phi^2} (\vec{\nabla}\cdot\vec{\vec{\sigma}})_1 ,
\end{split}
\end{equation}
and
\begin{equation}
\begin{split}
\sum_{j=1}^{3} q_{j2} \frac{d\theta_j}{dt}&=
\frac{d\theta_2}{dt}
-\frac{1}{2}\frac{d\theta_1}{dt}
-\frac{1}{2}\frac{d\theta_3}{dt}  \\ &=
\frac{3 B^x}{2}(3 \partial_2^2 u_2 + \partial_1^2 u_2 + 2 \partial_1 \partial_2 u_1)  \\
&=\frac{1}{2 \phi^2} (\vec{\nabla}\cdot\vec{\vec{\sigma}})_2 .
\end{split}
\end{equation}
Thus setting
\be
\sum \vec{q}_j \frac{d\theta_j}{dt} = 0
\label{eq:mecheqdyn}
\ee
ensures elastic equilibrium as defined by Eqs.~\eqref{eq:m1} and \eqref{eq:m2}
or Eqs.~\eqref{eq:enemecheq1} and \eqref{eq:enemecheq2}.  In Appendices A, B, and C
we derive the corresponding equations for the one-dimensional case, and for bcc and
fcc crystals in three dimensions.

	The goal here is to develop a method that incorporates
elasticity, dislocations, crystallization and all the features contained in
the PFC and related amplitude models that is also consistent with instantaneous
elastic equilibrium.  This can be achieved by solving
Eq. (\ref{eq:motions}) subject to condition given by Eq. (\ref{eq:mecheqdyn}).
To test our approach we have performed numerical calculations for some selected
systems, where we expect the mechanical equilibrium constraint to influence the dynamics.
A description of these calculations are given in the next section.

\section{Numerical tests}\label{sec:numerics}
	
	In this section we discuss the time evolution of the amplitudes using standard
conjugate gradient dynamics as described by Eq.~\eqref{aevolution}, and dynamics subject to the
elastic equilibrium condition Eq. ~\eqref{eq:mecheqdyn}. The corresponding 1D equations
are \eqref{eq:1ddynamics} and \eqref{eq:1Dequilibrium} in Appendix C.

Eq.~\eqref{aevolution} and its one dimensional counterpart \eqref{eq:1ddynamics} were solved
using a semi-implicit time stepping scheme, where the nonlinear terms of the
dynamical equations are treated explicitly while the linear terms are treated implicitly.
The spatial derivatives were calculated using fast
Fourier transforms. Evolution of the amplitudes according to Eq.~\eqref{aevolution} is
referred to as standard conjugate gradient dynamics while the other approach used is
time evolution with elastic equilibration. Numerically the elastic
equilibration is formulated as follows.

\subsection{Elastic equilibration}
	For a two dimensional triangular system it is straightforward to
show that Eq. \eqref{eq:mecheqdyn} can be written as
\be
-\frac{d\theta_i}{dt}+\frac{1}{2} \left(
\frac{d\theta_j}{dt} +\frac{d\theta_k}{dt}\right) = 0,
\ee
with $i$, $j$ and $k$ being different.  Applying the
chain rule shows that these time derivatives
can be written down as functional
derivatives of the energy defined by Eq.~\eqref{eq:2Denergy} as
\begin{equation}
\frac{d \theta_j}{d t} = - \Im{\left[ \frac{1}{\eta_j} \frac{\delta F}{\delta \eta^*}
\right]} =
- \frac{1}{2}\frac{\delta F}{\delta \theta_j} \phi_j^{-2} .
\label{eq:etaeneidentity}
\end{equation}
Thus the elastic equilibrium condition Eq. (\ref{eq:mecheqdyn}) can be written as
\begin{equation}
-\frac{\delta F}{\delta \theta_i} \phi_i^{-2} + \frac{1}{2}\left(
\frac{\delta F}{\delta \theta_j} \phi_j^{-2}
+\frac{\delta F}{\delta \theta_k} \phi_k^{-2}
\right) = 0.
\label{eq:mecheqen}
\end{equation}
The algorithm works as follows:
\begin{enumerate}
\item Set the initial configuration.
\item Equilibrate $\theta_i$ for all $i$ by solving Eq. \eqref{eq:mecheqen}.
\item Calculate time evolution using Eq.~\eqref{aevolution} for one step.
\item Go to 2.
\end{enumerate}

In 1D the elastic equilibration is simpler. We directly minimize the energy with
respect to $\theta$ i.e. the deformation field as
\begin{equation}
\frac{\delta F_{1D}}{\delta \theta}=0,
\end{equation}
The energy $F_{1D}$ is defined by Eq.~\eqref{eq:1Denergy}.

For numerical calculations the functional derivatives with respect to $\theta$
in both the 1D and the 2D cases are calculated by taking the imaginary part
of the conjugate gradient time evolution as suggested by Eq.~\eqref{eq:etaeneidentity}.

\subsection{Compression in one dimension}
\begin{figure}
\includegraphics[width=0.45\textwidth]{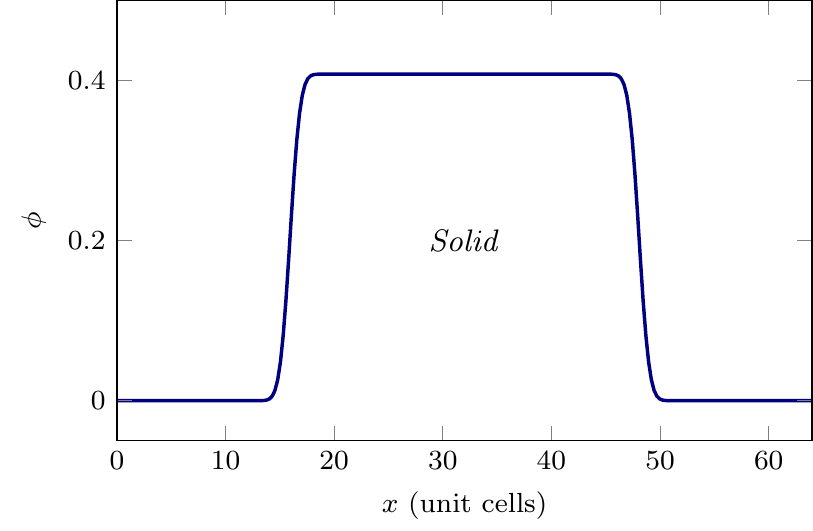}
\caption{Initial order parameter field: a solid block immersed in an undercool.}
\label{pic:phistart}
\end{figure}

\begin{figure}
\includegraphics[width=0.45\textwidth]{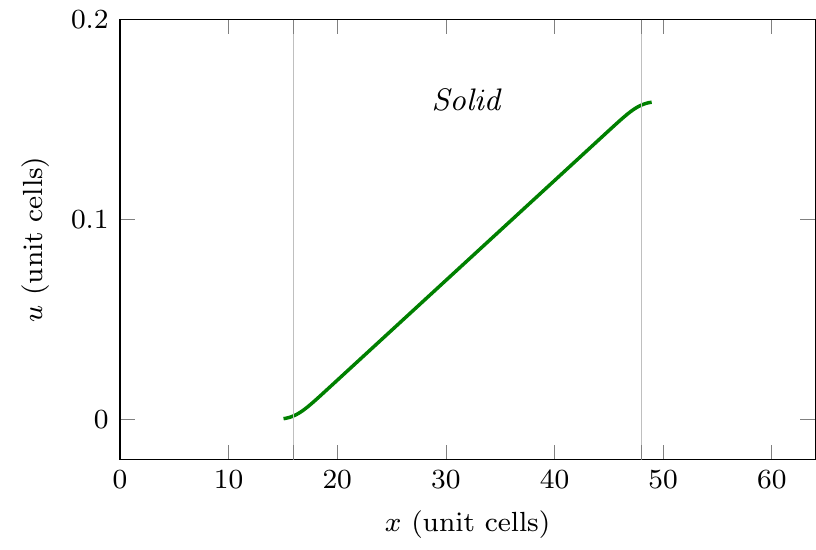}
\caption{Initial deformation field of the compressed 1D system.}
\label{pic:thetastart}
\end{figure}

Here we describe a test of the dynamics with elastic equilibrium imposed as
described in App.~\ref{sec:e1d}. We start with a compressed solid body immersed
in an undercooled liquid as seen in Figs.~\ref{pic:phistart} and
\ref{pic:thetastart} \footnote{For convenience we define the deformation field
as $\vec{q}_j \cdot \vec{u} = \theta_j$. For this reason our $\vec{u}$ has a
different sign than the real deformation field and the picture shows
compression instead of stretching.}. For the numerical work in this paper
we used the dimensionless parameters $\Delta B=-0.5$,
$B^x = 1$ and $v=1$. The size of the 1D system here is $256$ and the
spatial discretisation size is about $1.57$. We used a
time step of $0.05$ for the
evolution of the complex amplitudes.

\begin{figure}
\includegraphics[width=0.45\textwidth]{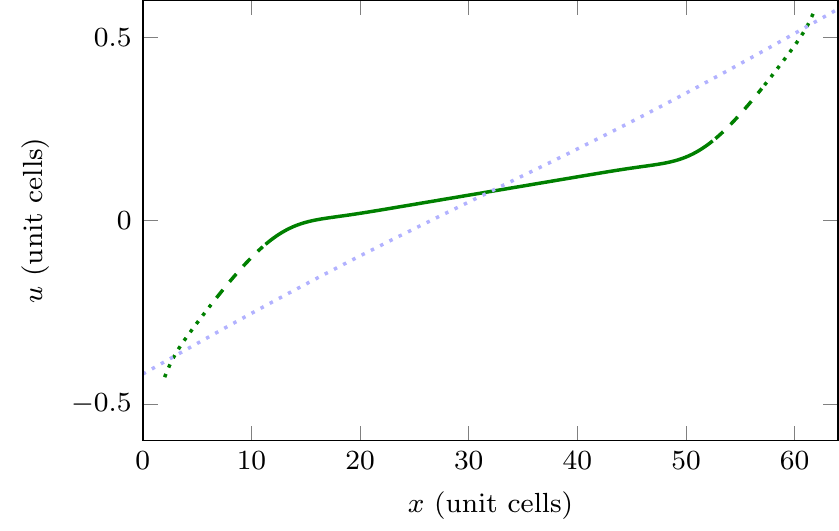}
\caption{The standard conjugate gradient method at times $10$, $20$ and $30$ as
shown by the solid, dashed and dotted lines, respectively.
The block solidifies entirely while the deformation field is left practically unchanged.
The blue dotted line shows the deformation field after $3000$ time units.
}
\label{pic:thetaout_all}
\end{figure}

The ground state of the system is a solid block with constant
$\phi \approx 0.408$. The evolution of $\phi$
is straightforward as it freezes towards the constant profile. What is
interesting is the evolution of the deformation field. Physical intuition tells
that the system should stretch very quickly, but what
actually happens with the standard conjugate gradient evolution is seen in
Fig.~\ref{pic:thetaout_all}. The system freezes too quickly for the elastic
instability to relax. When the system solidifies completely the elastic
stresses cannot relax any more since the periodic boundaries prevent any
stretching and the system remains in a strained state. It should be mentioned
that the deformation field cannot be defined in liquid and
therefore the domain of $u$ grows as the system solidifies.
\begin{figure}
\includegraphics[width=0.45\textwidth] {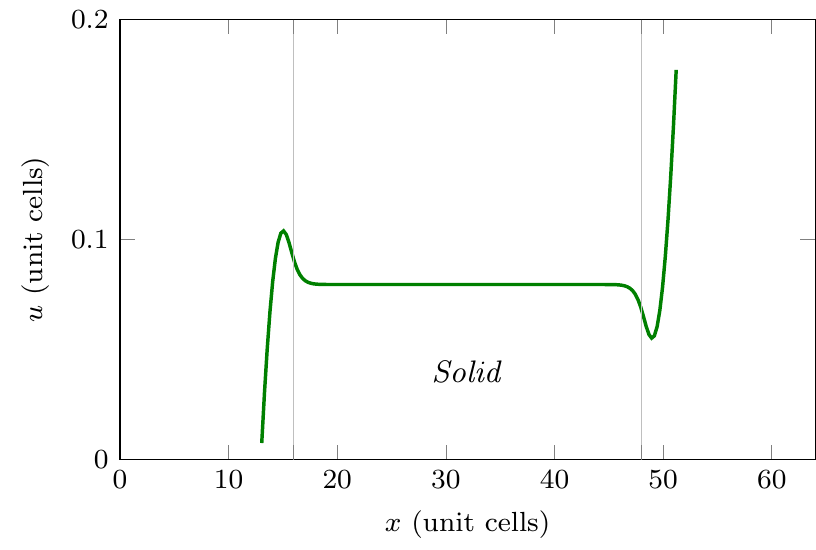}
\caption{Deformation field of the initial setting seen in
Fig.~\ref{pic:thetastart} after elastic equilibration.}
\label{pic:theta_eq}
\end{figure}

The elastic equilibration through the conjugate gradient dynamics is very slow.
It takes thousands of time units to get the equally strained
deformation profile in Fig.~\ref{pic:thetaout_all}
while it took only $39$ time units for the system to solidify.
The solidification process with the elastic equilibrium
imposed took only $19$ time units to solidify implying that even the dynamics of
the $\phi$ field are different depending on whether the dynamics is solved in
elastic equilibrium or not. The deformation field after the initial
equilibration is shown in Fig.~\ref{pic:theta_eq}.
The profile is simply stretched while the solid block grows.

\subsection{Grain rotation in two dimensions}
The elastic equilibration was also tested for the well known grain rotation
phenomena \cite{Wu:2012iu,Cahn04,Cahn06} in a two-dimensional triangular system.
In these simulations a circular grain is initially rotated by a certain
angle $\alpha$ creating dislocations
at the boundary between the circular grain and the surrounding solid body.

The classical description of grain boundary evolution states that the normal
velocity of the grain boundary is proportional to its curvature.
In the case of a circular grain the curvature can be written as
$R(t)^{-1}$, where $R(t)$ is the radius of the circle.
Now, $d/dt [R(t) ] \sim R(t)^{-1}$, which implies that $d/dt [R(t)^2]$ is a
constant. In other words, the area of the circular grain decreases linearly.
The shrinking in the normal direction of the boundary ensures that shrinking
of a circular grain is self similar, only the radius decreases.

Another consequence of the initial rotation is the rotation of the grain
while shrinking. This is due to the fact that for small rotation angles the
number of dislocations $n_D$ is conserved throughout the shrinking
(until a rapid final collapse of the grain) and is
proportional to the mismatch given by the rotation angle
$\alpha(t)$ times the grain boundary length $2\pi R(t)$ i.e.
$n_D \sim R(t) \alpha(t)$. This implies that $\alpha(t) \sim R(t)^{-1}$,
which makes the rotation angle grow as the grain radius shrinks.

To examine this phenomena we first conducted a set of simulations
with parameters identical to the ones chosen by Wu and Voorhees
\cite{Wu:2012iu} who examined grain rotation using the PFC model, i.e.,
Eq. (\ref{eq:pfc_evolution}).  A second set of simulations were also
conducted for parameters in which the difference between the standard
conjugate and the instantaneous elastic relaxation approaches is large.

The parametrisation of Wu and Voorhees \cite{Wu:2012iu} in our notation reads
$\Delta B = -0.014075$, $B^x = 1$, $\tau=0.585$ and $v=1$ and is from now on referred to
as the warm case (parametrization). We calculated the dynamics also with a
colder effective temperature by dropping the value of $\Delta B$ to $-0.05$
(cold parametrization). In both of these cases the initial rotation angle
was chosen to be 5° corresponding to calculations done in \cite{Wu:2012iu}.
All the calculations were performed using isotropic spatial discretisation of
$4.0$ and a time step of $1.0$. A simulation box of 1568$\times$1568 with periodic boundaries
was used for all calculations. This comprises about 216$\times$216 atoms of which the rotated
grain occupies about 1/4 with a diameter of 100 atoms.
\begin{figure}
\center
\subfigure[]
{\includegraphics[width=2.7cm]{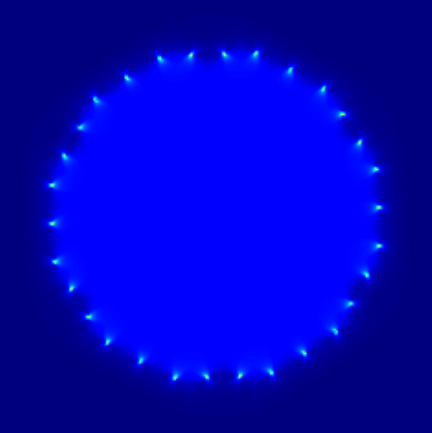}}\;
\subfigure[]
{\includegraphics[width=2.7cm]{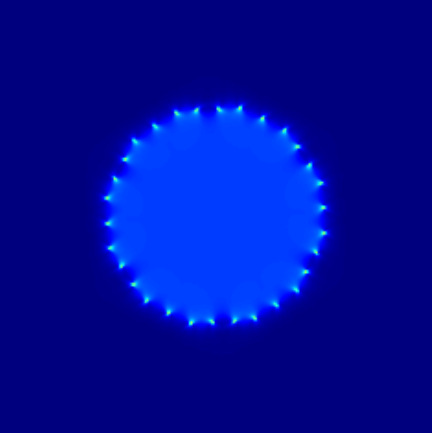}}\;
\subfigure[]
{\includegraphics[width=2.7cm]{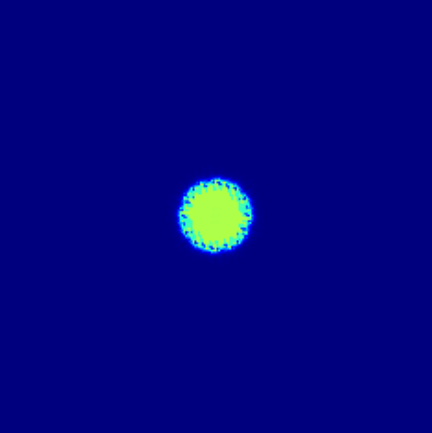}}
\caption{The norm of the gradient of the deformation field $\norm{\nabla \vec{u}}$.
Panels (a)-(c) show the time evolutions for warm parametrisation of Wu et al.
\cite{Wu:2012iu} at times
$10000$, $400000$, and $680000$, respectively.
Brighter color stands for greater value of the norm. The dots
on the perimeter show the dislocations at the grain boundary.
}
\label{pic:unorm}
\end{figure}

\begin{figure}
\includegraphics[width=0.45\textwidth] {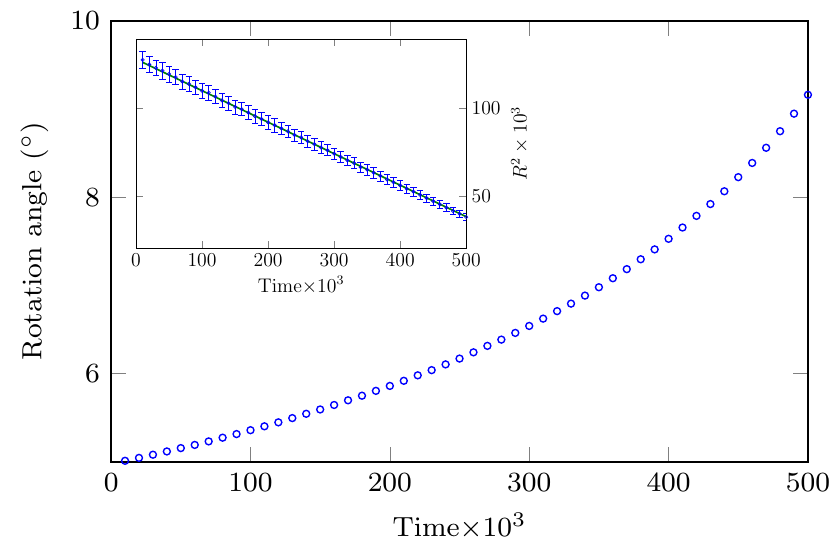}
\caption{Angle of the shrinking grain for the warm parametrisation.
Inset shows the corresponding squared radius.
}
\label{pic:angles_radii}
\end{figure}

Fig.~\ref{pic:unorm} shows the gradient of the deformation field
$\norm{\nabla \vec{u}}= \sqrt{\sum_{i,j} (\partial_i u_j)^2}$ without the
equilibration. For small angles this is proportional to the rotation angle
inside the circle since the deformation is a pure rotation.
The brighter colours at the boundary show the dislocations that join in the
last panel to vanish shortly afterwards. It must be noted that the grain
is shown in the original coordinates without any displacement.
The radius and the angle as a function of time for the warm
parametrisation can be seen in Fig.~\ref{pic:angles_radii}.
These results agree very well with those in Ref. \cite{Wu:2012iu} and
show that for this set of parameters the amplitude representation,
i.e., Eq. (\ref{aevolution}) accurately reproduces the full PFC
results, i.e., Eq. (\ref{eq:pfc_evolution}).

For the warm case the dynamics with elastic equilibration is indistinguishable
within the errors from the standard conjugate gradient dynamics. This is due to the
fact that the parameters were chosen very close to the liquid state to
avoid getting stuck at local energy minima. The elastic energies are very small
close to the liquid state and the equilibration does not make any
discernible difference.
A linear fit to the squared radius data gives a slope of $-0.181 \pm 0.01 $
in dimensionless units.

\begin{figure}
\includegraphics[width=0.45\textwidth] {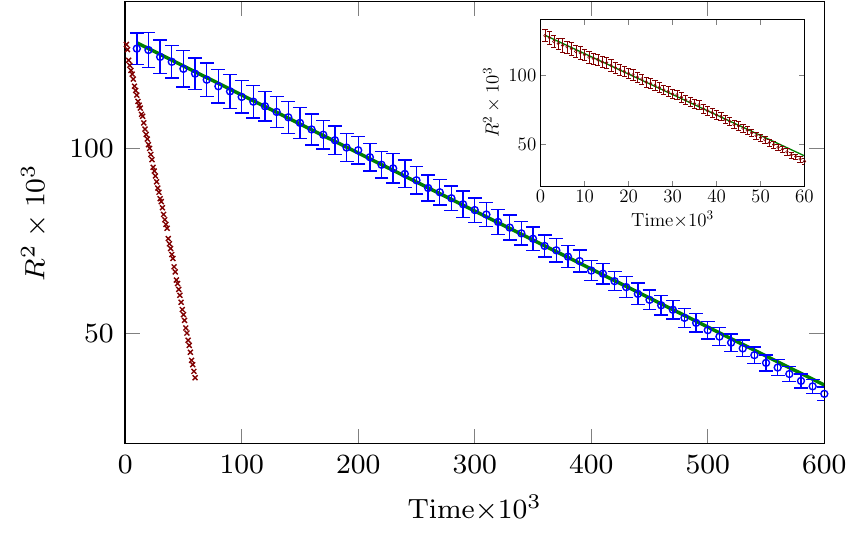}
\caption{The steep line with the red crosses shows the squared radius data
for the cold parametrisation with the elastic equilibration while the circles
show the equivalent data for the standard conjugate gradient dynamics.
The inset shows the data for the equilibrated dynamics with error bars.
}
\label{pic:radii_cold_all}
\end{figure}

The situation is very different in the cold case. Linear scaling of the squared
radius still holds for both the equilibrated and the standard conjugate gradient
dynamics, but the time scales are completely different as seen
in Fig.~\ref{pic:radii_cold_all}. The slope of the linear fit to the squared
radius data is $-0.157 \pm 0.01$ for the cold case with conjugate gradient
dynamics implying that the dynamics is slightly faster with the
warm parametrisation as expected. The corresponding slope of the linear fit
for the equilibrated dynamics in the cold case is $-1.47 \pm 0.1$
that is almost ten times faster than the slope with the conjugate gradient dynamics.
This suggests that with the standard conjugate gradient dynamics the inability
to quickly reach elastic equilibrium severely hinders shrinkage.

\section{Summary and Conclusions}\label{sec:summary}
We have proposed a method to separately relax elastic excitations
in the amplitude expansion picture of the PFC model during non-conserved
dissipative dynamics. This approach is shown to be consistent in the small deformation
limit with the theory of linear elasticity. The numerical tests suggest that
the approach indeed relaxes elastic excitations and furthermore that this relaxation
can considerably change the dynamics.

An interesting result of the test cases is that the elastic excitations are
extremely resilient in the diffusive time scales. It seems that the system
tends to prefer solidification to relaxing simple elastic stretches and
strains. This is unfortunate especially with systems driven out of equilibrium.
Traditional conjugate gradient dynamics allows only for diffusive
transportation of the information of any strains in the solid body.
This is in conflict with the theory of elasticity that predicts
ballistic transport of small displacements.
This problem is present both in the PFC dynamics
and the conjugate gradient dynamics of the amplitude expansion model,
and becomes even more important when the system is mechanically driven out of equilibrium. Our
approach to equilibrate the elastic excitations remedies this
inherent shortcoming in the standard diffusive dynamics.

\begin{acknowledgments}
This work has been supported in part by the Academy of
Finland through its COMP CoE grant no. 251748.
K.R.E. wants to thank the Aalto Science Institute for a Visiting
Professorship grant and
acknowledges support from NSF Grant No. DMR-0906676.
\end{acknowledgments}

\appendix

\section{Elastic excitations in bcc crystals}
\label{sec:BCC}

\subsection{Elastic equilibrium from energy}

To describe the bcc lattice first mode approximation we use the following
reciprocal lattice vectors \begin{equation*}
\begin{split}
&\vec{q}_1=(1,1,0)/\sqrt{2}, \\
&\vec{q}_2=(1,0,1)/\sqrt{2},  \\
&\vec{q}_3=(0,1,1)/\sqrt{2},  \\
&\vec{q}_4=(0,1,-1)/\sqrt{2}, \\
&\vec{q}_5=(1,-1,0)/\sqrt{2}, \\
&\vec{q}_6=(-1,0,1)/\sqrt{2}. \\
\end{split}
\end{equation*}

The complete energy for a 3D bcc system is written down in \cite{Elder10}.
As for the 2D case, the only term giving rise to elastic energy in
the energy is again the term
\begin{equation}
4 B^x \phi^2 \sum_{k,l} q_{jk}q_{jl} (\partial_k \theta_j) (\partial_l \theta_j),
\end{equation}
and the elastic part of the energy can be defined as
\begin{equation}
F_{el} = \int_{\Omega} d\vec{r}
\left[ 4 B^x \phi^2 \sum_{k,l,j} q_{jk}q_{jl} (\partial_k \theta_j) (\partial_l \theta_j) \right].
\end{equation}
This in terms of the linear strain tensor is
\begin{equation}
\begin{split}
F_{el} &= \int_{\Omega} d\vec{r} \left[ 8 B^x \phi^2 \left( 2 \sum_{i=1}^3 \varepsilon_{ii}^2
+4 \varepsilon_{12}^2 + 4 \varepsilon_{13}^2
\right.\right. \\ & \left. \left.
+ 4\varepsilon_{23}^2
+2 \varepsilon_{22} \varepsilon_{33} +2 \varepsilon_{11}
\varepsilon_{22}+2 \varepsilon_{11} \varepsilon_{33}
\vphantom{\sum_{i=1}^3}
\right) \right].
\end{split}
\end{equation}
Now the stress tensor becomes
\begin{equation}
\begin{split}
\sigma_{ii} &= \frac{\delta F_{el}}{\delta \varepsilon_{ii}} \\&
= 32 B^x \phi^2 \varepsilon_{ii}
+ 16 B^x \phi^2 \varepsilon_{jj} + 16 B^x \phi^2 \varepsilon_{kk},
\end{split}
\end{equation}
for $i$, $j$, $k$ different and
\begin{equation}
\sigma_{ij} = \frac{1}{2} \frac{\delta F_{el}}{\delta \varepsilon_{ij}} =
32 B^x \phi^2 \varepsilon_{ij},
\end{equation}
for $i \neq j$.
Writing the elastic equilibrium $\sum_j \partial_j \sigma_{ij} =0$
in terms of the components of $\vec{u}$ gives
\begin{equation}
\label{meceq}
2 \partial_i^2 u_i + (\partial_j^2 + \partial_k^2) u_i + 2
\partial_i (\partial_j u_j + \partial_k u_k) =0,
\end{equation}
for all $i$, $j$, $k$ different.

The elastic constants of the cubic crystal symmetry can be obtained as
\begin{align}
\sigma_{ii} &= C_{11} \varepsilon_{ii} + C_{12} (\varepsilon_{jj} + \varepsilon_{kk}),
\quad \textrm{$i$, $j$, $k$ different;} \\
\sigma_{ij} &= 2 C_{44} \varepsilon_{ij},\quad \textrm{$i \neq j$}
\end{align}
giving $C_{11}=32 B^x \phi^2$, $C_{12}=16 B^x \phi^2$ and $C_{44}=16 B^x \phi^2$.

\subsection{Elastic equilibrium from dynamics}

The equivalent of Eq.~\eqref{aevolution} can be found from \cite{Elder10}.
The dynamical equations of motion for $\phi$ are
\be
\label{eq2}
\vq_j\cdot\nabla\nabla^2\theta_j=c_j,
\ee
where $c_j$ are constants. We will now combine different components of Eq.~(\ref{eq2}):
$(j=1)+(j=5)$, $(j=2)+(j=6)$, $(j=1)-(j=5)$, $(j=2)-(j=6)$, $(j=3)-(j=4)$
and finally $(j=3)+(j=4)$, respectively, yield the following relations,
\begin{equation}
\label{eq3}
\begin{split}
&
(\px^3 + \px\py^2 +\px\pz^2)u_1  \\
+& (\py\px^2+\py\pz^2+\py^3)u_2=d_1,
\end{split}
\end{equation}
\begin{equation}
\label{eq4}
\begin{split}
&
(\px^3 + \px\py^2 + \px\pz^2)u_1\\+&
 (\pz\px^2 + \pz\py^2 + \pz^3)u_3=d_2,
\end{split}
\end{equation}
\begin{equation}
\label{eq5}
\begin{split}
&
(\px^3 +\px\py^2 +\px\pz^2)u_2 \\+&
(\py\px^2 +\py\pz^2 + \py^3)u_1=d_3,
\end{split}
\end{equation}
\begin{equation}
\label{eq6}
\begin{split}
&
(\px^3 + \px\py^2 + \px\pz^2)u_3\\+&
(\pz\px^2 +\pz\py^2 + \pz^3)u_1=d_4,
\end{split}
\end{equation}
\begin{equation}
\label{eq7}
\begin{split}
&
(\py^3 +\py\px^2 +\py\pz^2)u_3\\+&
(\pz\px^2 +\pz\py^2 +\pz^3)u_2=d_5,
\end{split}
\end{equation}
\begin{equation}
\label{eq8}
\begin{split}
&
(\py^3 +\py\px^2 +\py\pz^2)u_2\\+&
(\pz\px^2 +\pz\py^2 +\pz^3)u_3=d_6,
\end{split}
\end{equation}
where $d_j$ are constants. Now, $\frac{1}{2}\px$[Eq.~(\ref{eq3})+Eq.~(\ref{eq4})]
$-\frac{3}{2}\px$[Eq.~(\ref{eq8})] $+\py$[Eq.~(\ref{eq5})] $+\pz$[Eq.~(\ref{eq6})] gives
\be
\label{eq9}
\nabla^4u_1=0.
\ee
Furthermore, $\px$[Eq.~(\ref{eq5})] $+\pz$[Eq.~(\ref{eq7})] $+\frac{1}{2}\py$
[Eq.~(\ref{eq3}) + Eq.~(\ref{eq8})] $-\frac{3}{2}\py$[Eq.~(\ref{eq4})] yields
\be
\label{eq10}
\nabla^4u_2=0.
\ee
Finally, from $\px$[Eq.~(\ref{eq6})] $+\py$[Eq.~(\ref{eq7})] $-\frac{3}{2}\pz$
[Eq.~(\ref{eq3})] $+\frac{1}{2}\pz$[Eq.~(\ref{eq4}) + Eq.~(\ref{eq8})], it follows that
\be
\label{eq11}
\nabla^4u_3=0.
\ee
Thus, $\nabla^4\theta_j=0$.

The time evolution for the $\theta_j$ fields can be written as
\begin{equation}
\frac{d\theta_j}{dt} = -B^x \nabla^4\theta_j+ 4 B^x \left(\vq_j\cdot\nabla\right)^2\theta_j,
\end{equation}
which now simplifies into
\begin{equation}
\label{eq:thetadynbcc}
\frac{d\theta_j}{dt} = 4 B^x \left(\vq_j\cdot\nabla\right)^2\theta_j =
4 B^x \left(\vq_j\cdot\nabla\right)^2 \vec{q}_j \cdot \vec{u}.
\end{equation}
The elastic equilibrium equations are written down with the help of Eq.~\eqref{eq:mecheqdyn} as
\begin{equation}
\sum_{j=1}^6 \vec{q}_j \frac{d \theta_j}{dt} = 0.
\end{equation}
This becomes
\begin{align}
\label{eq:bcc1}
\frac{d\theta_1}{dt}+\frac{d\theta_2}{dt}+\frac{d\theta_5}{dt}-\frac{d\theta_6}{dt}&=0; \\
\label{eq:bcc2}
\frac{d\theta_1}{dt}+\frac{d\theta_3}{dt}+\frac{d\theta_4}{dt}-\frac{d\theta_5}{dt}&=0; \\
\label{eq:bcc3}
\frac{d\theta_2}{dt}+\frac{d\theta_3}{dt}-\frac{d\theta_4}{dt}+\frac{d\theta_6}{dt}&=0.
\end{align}
The time derivatives can be replaced using Eq.~\eqref{eq:thetadynbcc} giving
\begin{align}
\label{eq13}
2\px^2u_1+\py^2u_1+\pz^2u_1+2\px\py u_2+2\px\pz u_3&=0; \\
\label{eq14}
2\py^2u_2+\px^2u_2+\pz^2u_2+2\px\py u_1+2\py\pz u_3&=0; \\
\label{eq15}
2\pz^2u_3+\px^2u_3+\py^2u_3+2\px\pz u_1+2\py\pz u_2&=0,
\end{align}
which give $\left(\nabla\cdot\sigma\right)_i=0$ for $i=1,2,3$ respectively.

\section{Elastic excitations in fcc crystals}
\label{sec:FCC}
In order to reproduce the fcc lattice symmetry, two different sets of reciprocal lattice
vectors of different scales are needed (two-mode approximation) that are both cubically
symmetric. Let us choose them to be
\begin{equation*}
\begin{split}
&\vec{q}_1=(-1,1,1)/\sqrt{3}, \\
&\vec{q}_2=(1,-1,1)/\sqrt{3},  \\
&\vec{q}_3=(1,1,-1)/\sqrt{3},  \\
&\vec{q}_4=(-1,-1,-1)/\sqrt{3}, \\
&\vec{q}_5=2(0,0,1)/\sqrt{3}, \\
&\vec{q}_6=2(1,0,0)/\sqrt{3},  \\
&\vec{q}_7=2(0,1,0)/\sqrt{3}. \\
\end{split}
\end{equation*}

\subsection{Elastic equilibrium from the energy}

Full energy for the fcc system can be found from Ref. \cite{Elder10}.
Again, the only term giving rise to elastic energy in the energy is the term
\begin{equation}
4 B^x \phi^2 \sum_{k,l} q_{jk}q_{jl} (\partial_k \theta_j) (\partial_l \theta_j),
\end{equation}
and the elastic part of the energy can be defined as
\begin{equation}
F_{el} = \int_{\Omega} d\vec{r} \left[ 4 B^x \phi^2 \sum_{k,l,j}
q_{jk}q_{jl} (\partial_k \theta_j) (\partial_l \theta_j) \right].
\end{equation}
This in terms of the linear strain tensor is
\begin{equation}
\begin{split}
F_{el} &= \int_{\Omega} d\vec{r} \left[ 16 B^x \phi^2 \left( 5 \sum_{i=1}^3 \varepsilon_{ii}^2
+4 \varepsilon_{12}^2 + 4 \varepsilon_{13}^2
\right. \right. \\& \left. \left.
+ 4\varepsilon_{23}^2
+2 \varepsilon_{22} \varepsilon_{33} +2 \varepsilon_{11}
\varepsilon_{22}+2 \varepsilon_{11} \varepsilon_{33}
\vphantom{\sum_{i=1}^3}
\right) \right].
\end{split}
\end{equation}
Now the stress tensor becomes
\begin{equation}
\begin{split}
\sigma_{ii} &= \frac{\delta F_{el}}{\delta \varepsilon_{ii}} \\
&= 160 B^x \phi^2 \varepsilon_{ii}
+ 32 B^x \phi^2 \varepsilon_{jj} + 32 B^x \phi^2 \varepsilon_{kk},
\end{split}
\end{equation}
for all $i$, $j$, $k$ different and
\begin{equation}
\sigma_{ij} = \frac{1}{2} \frac{\delta F_{el}}{\delta
\varepsilon_{ij}} = 64 B^x \phi^2 \varepsilon_{ij},
\end{equation}
for all $i \neq j$.
Writing the elastic equilibrium $\sum_j \partial_j \sigma_{ij} =0$
in terms of the components of $\vec{u}$ gives
\begin{equation}
5 \partial_i^2 u_i + (\partial_j^2 + \partial_k^2 )u_i +
2 \partial_i (\partial_j u_j + \partial_k u_k) =0,
\end{equation}
for all $i$, $j$, $k$ different.

The elastic constants are again from the cubic crystal symmetry
\begin{align}
\sigma_{ii} &= C_{11} \varepsilon_{ii} + C_{12} (\varepsilon_{jj} +
\varepsilon_{kk}), \quad \textrm{$i$, $j$, $k$ different}; \\
\sigma_{ij} &= 2 C_{44} \varepsilon_{ij},\quad \textrm{$i \neq j$}.
\end{align}
Now $C_{11}=160 B^x \phi^2$, $C_{12}=32 B^x \phi^2$ and $C_{44}=32 B^x \phi^2$.

\subsection{Elastic equilibrium from dynamics}

The evolution of the complex amplitudes can be found from Ref. \cite{Elder10}.
When making again the assumption that $\eta_j=\phi_j e^{i \vec{q}_j \cdot \vec{u}}$
with constant $\phi_j$ and going to linear order in $\vec{u}$ gives us equation for
the amplitudes as
\begin{equation}
\sum_{k,l} q_{jk} \partial_k v_l q_{jl}=C_1,
\label{phi1}
\end{equation}
for $j=1,2,3,4$ and
\begin{equation}
\sum_{k,l} q_{jk} \partial_k v_l q_{jl}=C_2,
\label{phi2}
\end{equation}
for $j=5,6,7$. Here $C_1$ and $C_2$ are constants consisting of constant
amplitudes and model parameters and $\vec{v}=\nabla^2 \vec{u}$.
Again we need to show that the biharmonic equation $\nabla^4 \vec{u}$ follows.
Inserting the reciprocal vectors $\vec{q}_5$, $\vec{q}_6$, and
$\vec{q}_7$ in Eq.~\eqref{phi2} gives
\begin{equation}
\partial_i v_i = C_2,
\label{eq:fccdiv}
\end{equation}
for all $i$.

Next, let us open Eq.~\eqref{phi1}
\begin{equation}
\begin{split}
&q_{j1}^2 \partial_1 v_1 + q_{j2}^2 \partial_2 v_2 + q_{j3}^2 \partial_3 v_3 \\ &
+ q_{j1}q_{j2} (\partial_1 v_2 + \partial_2 v_1)
+ q_{j1}q_{j3} (\partial_1 v_3 + \partial_3 v_1) \\&
+q_{j2}q_{j3} (\partial_2 v_3 + \partial_3 v_2)=C_2.
\end{split}
\label{eq:biharmonics}
\end{equation}
We can take advatage of the fact that the set $\{ \vec{q}_j\}$ is invariant
under cubic symmetry operations. Using reflections of the coordinate $i$
\footnote{$\vec{q}_{ji} \to -\vec{q}_{ji}$ and $q_{jk}$ stays invariant when
$k\neq i$.} and subtracting from both sides of \eqref{eq:biharmonics} it follows that
\begin{equation}
q_{ji}q_{jk} (\partial_i v_k + \partial_k v_i)+q_{ji}q_{jl}
(\partial_i v_l + \partial_l v_i)=0,
\end{equation}
or
\begin{equation}
q_{jk} (\partial_i v_k + \partial_k v_i)+q_{jl} (\partial_i v_l + \partial_l v_i)=0,
\label{eqfccq}
\end{equation}
since $q_{ji}$ are non-zero for $i=1,2,3,4$.
Here the indices $i$, $k$, $l$ are all different so Eq. \eqref{eqfccq} applies
for all the permutations of $1$, $2$ and $3$.
Using reflection on $q_{jl}$ and adding to Eq.~\eqref{eqfccq} gives
\begin{equation}
q_{jk} (\partial_i v_k + \partial_k v_i)=0,
\end{equation}
or
\begin{equation}
(\partial_i v_k + \partial_k v_i)=0,
\end{equation}
for all $k\neq i$. Taking derivative $\partial_i$ it follows that
\begin{equation}
\partial_i^2 v_k = 0,
\end{equation}
since $\partial_k \partial_i v_i=0$ according to Eq.~\eqref{eq:fccdiv}. Now
\begin{equation}
(\partial_1^2+\partial_2^2+\partial_3^2) v_i=0,
\end{equation}
for all $i$ i.e. $\nabla^4 \vec{u}=0$.

The time evolution for $\theta_j$ after applying the biharmonic equation becomes
\begin{equation}
\frac{d\theta_j}{dt} = 4 B^x (\vec{q}_j \cdot \vec{\nabla})^2 \vec{q}_j \cdot \vec{u},
\label{theta1}
\end{equation}
or in terms of components
\begin{equation}
\label{eq:temp1}
\frac{d\theta_j}{dt} = 4 B^x \sum_{k,l,p} q_{jk} q_{jl} \partial_k \partial_l q_{jp} u_p,
\end{equation}
for all $i=1-7$.

The elastic equilibrium condition can be written down using Eq.~\eqref{eq:mecheqdyn} as
\begin{equation}
\sum_{j=1}^7 \vec{q}_j \frac{d \theta_j}{dt} = 0,
\end{equation}
giving
\begin{align}
\label{eq:fcc1}
-\frac{d\theta_1}{dt}+\frac{d\theta_2}{dt}+\frac{d\theta_3}{dt}-\frac{d\theta_4}{dt}
+ 2\frac{d\theta_6}{dt}&=0; \\
\label{eq:fcc2}
\frac{d\theta_1}{dt}-\frac{d\theta_2}{dt}+\frac{d\theta_3}{dt}-\frac{d\theta_4}{dt}
+ 2\frac{d\theta_7}{dt}&=0; \\
\label{eq:fcc3}
\frac{d\theta_1}{dt}+\frac{d\theta_2}{dt}-\frac{d\theta_3}{dt}-\frac{d\theta_4}{dt}
+ 2\frac{d\theta_5}{dt}&=0.
\end{align}
Using Eq.~\eqref{eq:temp1} this gives
\begin{align}
5 \partial_1^2 u_1 + (\partial_2^2 + \partial_3^2)u_1 +
2 \partial_1 (\partial_2 u_2 + \partial_3 u_3) &= 0; \\
5 \partial_2^2 u_2 + (\partial_1^2 + \partial_3^2)u_2 +
2 \partial_2 (\partial_1 u_1 + \partial_3 u_3) &= 0; \\
5 \partial_3^2 u_3 + (\partial_1^2 + \partial_2^2)u_3 +
2 \partial_3 (\partial_1 u_1 + \partial_2 u_2) &= 0,
\end{align}
which constitutes to $(\nabla \cdot \sigma)_i = 0$  for indices $i=1,2,3$,
respectively i.e. gives the elastic equilibrium condition.

\section{Elastic excitations in 1D}\label{sec:e1d}

\subsection{Separation of complex amplitudes}

Energy for a 1D system can be written as
\begin{equation}
F_{1D} = \int_{\Omega} dx [\Delta B |\eta |^2 + B^x |\mathcal{G} \eta |^2 +
\frac{3}{2}v |\eta |^4],
\label{eq:1Denergy}
\end{equation}
where $\mathcal{G}=\partial_x^2 + 2i \partial_x$. The time evolution for $\eta$ is
\begin{equation}
\begin{split}
\frac{d \eta}{d t} &= - \frac{\delta E_{1D}}{\delta \eta ^*} =
\frac{d \phi}{d t} e^{i \theta}
+ i \phi \frac{d \theta}{d t} e^{i \theta} \\
&= -\left\lbrace
\Delta B \eta + B^x \mathcal{G}^2\eta + 3v |\eta|^2 \eta
\right\rbrace.
\end{split}
\label{eq:1ddynamics}
\end{equation}
Opening the right-hand side and separating the complex and the real parts gives
\begin{equation}
\begin{split}
\frac{d \phi}{d t} &=
-\Delta B \phi \\&
- B^x [4 \phi (\partial_x \theta)^2
+4 \phi (\partial_x \theta)^3
+\phi (\partial_x \theta)^4 \\&
 -12 (\partial_x\phi) (\partial_x^2 \theta)
 - 12 (\partial_x \theta) (\partial_x \phi) (\partial_x^2 \theta) \\&
-3\phi (\partial_x^2 \theta)^2
-4 \partial_x^2 \phi
- 12 (\partial_x \theta) (\partial_x^2 \phi)  \\&
- 6 (\partial_x \theta)^2 (\partial_x^2 \phi)
-4 \phi (\partial_x^3 \theta ) \\&
 -4 \phi (\partial_x \theta) (\partial_x^3 \theta)
+ \partial_x^4 \phi
] - 3v \phi ^3,
\end{split}
\label{eq:1Dphievolution}
\end{equation}
and
\begin{equation}
\begin{split}
\phi \frac{d \theta}{d t} &=-
B^x[
-8 (\partial_x \theta ) (\partial_x \phi)
- 12 (\partial_x \theta)^2 (\partial_x \phi) \\&
-4 (\partial_x \theta)^3 (\partial_x \phi )
-4 (\partial_x^2 \theta) \phi \\&
 -12 \phi (\partial_x \theta) (\partial_x^2 \theta)
 -6 (\partial_x \theta)^2 (\partial_x^2 \theta) \phi \\&
+6 (\partial_x^2 \theta) (\partial_x^2 \phi)
+4 (\partial_x^3 \theta) (\partial_x \phi) \\&
 +4 (\partial_x^3 \phi)
 + 4 (\partial_x \theta) (\partial_x^3 \phi)
 +(\partial_x^4 \theta) \phi
].
\end{split}
\label{eq:1Dthetaevolution}
\end{equation}

In one dimension the deformation field $u=\theta$,
which gives an expression for the elastic equilibrium condition
\begin{equation}
\frac{d \theta}{dt}=0.
\label{eq:1Dequilibrium}
\end{equation}

\subsection{Linear elasticity}
Let us consider evolution of $\phi$ given by Eq. \eqref{eq:1Dphievolution}.
Writing $\theta (x) = u(x)$ and going to linear order in the deformation field $u$
gives an equation for constant $\phi$
\begin{equation}
\partial_x^3 u = \frac{\Delta B+ 3v \phi^2}{4 B^x} = C,
\end{equation}
where $C=\textit{const.}$, from which it follows that
\begin{equation}
\partial_x^4 u = 0.
\end{equation}

Writing down the condition of Eq. \eqref{eq:1Dequilibrium} for
Eq.~\eqref{eq:1Dthetaevolution} in the linear regime gives
\begin{equation}
-4 \partial_x^2 u + \partial_x^4 u = 0,
\end{equation}
implying that
\begin{equation}
\partial_x^2 u =0.
\end{equation}

The same relation follows in the linear elasticity limit from
the energy by taking the functional derivative with respect to the
deformation field i.e. demanding that
\begin{equation}
\frac{\delta F_{1D}}{\delta u} = \frac{\delta F_{1D}}{\delta \theta} = 0.
\end{equation}

\end{document}